\documentclass[aps,twocolumn,showpacs]{revtex4}
\usepackage{epsfig,amsmath}
\begin{document}
\def\I.#1{\it #1}
\def\B.#1{{\bf #1}}
\def\C.#1{{\cal  #1}}
\title{Inverse Cascade Regime in Shell Models of 2-Dimensional
Turbulence}
\author{Thomas Gilbert}
\email{thomas.gilbert@weizmann.ac.il}
\author{Victor S. L'vov}
\email{victor.lvov@weizmann.ac.il}
\homepage{http://lvov.weizmann.ac.il}
\author{Anna Pomyalov}
\email{anna.pomyalov@weizmann.ac.il}
\author{Itamar Procaccia}
\email{itamar.procaccia@weizmann.ac.il}
\homepage{http://www.weizmann.ac.il/chemphys/cfprocac}   
\affiliation{Department of~~Chemical Physics, The Weizmann Institute of
   Science, Rehovot 76100, Israel}
%
%
\begin{abstract}
We consider shell models that display an inverse energy cascade
similar to 2-dimensional turbulence (together with a direct 
cascade of an enstrophy-like invariant).  Previous attempts
to construct such models ended negatively, stating that shell models 
give rise to a ``quasi-equilibrium" situation with equipartition of the
energy among the shells. 
We show analytically that the quasi-equilibrium state predicts
its own disappearance upon changing the model parameters in favor of
the establishment of an inverse cascade regime with K41 scaling. 
The latter regime
is found where predicted, offering a useful model to study inverse cascades.

\end{abstract}
   \maketitle
The inverse energy cascade in 2-dimensional Navier-Stokes turbulence
is an important phenomenon with implications for geophysical flows \cite{67Kra}.
In addition, it had been found that correlation functions and
structure functions obey very
closely Kolmogorov scaling (so-called K41), with only minute anomalous
corrections, in contradistinction to 3-dimensional turbulence in which intermittency
corrections to K41 scaling are sizable \cite{01Tab}. This difference is well documented
\cite{93SY,98PT,00BCV} but not yet understood.
It is therefore tempting to construct simple models of the phenomenon.
Indeed, several attempts were made to construct shell models for this
purpose, \cite{94ABCFPV,96DM}.
So far these attempts ended negatively, failing to find a statistical
steady state in which 
energy flows from smaller to larger scales {\em together} with having a 
Kolmogorov energy spectrum. Rather, it was thought that 
whenever energy flew ``backwards", the 
statistical steady state settled close to thermodynamic equilibrium. In this
letter we show that there actually exists a wide range of parameter values
for which 
shell models display the wanted behavior, thereby offering useful testing
grounds for ideas on 2-dimensional turbulence.

We discuss the issue in the framework of the Sabra shell model
\cite{98LPPPV}. Like all shell models \cite{98BJPV} this represents a truncated Fourier  
representation of the Navier-Stokes equations. The Sabra model reads
\begin{eqnarray}\label{sabra} 
\frac{d u_n}{dt} &=& i(a k_{n+1}u_{n+1}^*u_{n+2} + b k_n
 u_{n-1}^*u_{n+1}
\nonumber\\&-&c k_n u_{n-1}u_{n-2})- \gamma_n u_n + f_n  \ ,
\end{eqnarray}
where  the dissipative term $\gamma_n$ reads $\nu  k_n^{2 \alpha} 
+ \mu k_n^{-2 \beta}$, with
$\nu$ and $\mu$ being the viscosity and drag coefficients respectively.
Here $u_n$
are complex numbers standing for the Fourier components of the velocity field
belonging to shell $n$, associated with wavenumbers $k_n$. The latter are
restricted to the set $k_n = k_0 \lambda^n$, with $\lambda$ being the 
spacing parameter, taken below to be 2. The forcing
$f_n$ is chosen here to act at intermediate values of $n$, $n=n_f$, allowing in
principle to study direct as well as inverse fluxes. The forcing is taken
random with Gaussian time correlations as in \cite{98LPPPV}; the amplitude of the 
forcing is fixed below to $1/\sqrt{2}$ in all cases. 
The dissipative terms $\gamma_n$ act
both on the smallest and the largest scales with their respective
(hyper)-viscosity and drag
exponents $\alpha$ and $\beta$; below we use $\alpha=\beta=2$.
 The dissipative terms become
dominant at the viscous and drag scales $n_d$ and $n_L$ respectively. We will always have
$n_L\ll n_f\ll n_d$. The coefficients
$a$, $b$ and $c$ are  adjustable parameters, with the constraint
$a + b + c = 0$ ensuring the conservation of energy in the dissipationless limit.
Choosing $a=1$ we explore the problem in terms of the single parameter $b$,
with $-2<b<0$ \cite{98LPPPV}.

It was shown before \cite{73Gle,88YO} that for $b<-1$ there exist two positive definite
invariants, the energy $E$ and the ``enstrophy" $H$,
\begin{equation}
E=\frac{1}{2}\sum_{n=1}^N |u_n|^2 \ , \quad H= \frac{1}{2}\sum_{n=1}^N 
\Big(\frac{-1}{b+1}\Big)^n|u_n|^2 \ ,
\end{equation}
which, in this case, are associated
with an inverse and direct fluxes respectively \cite{94ABCFPV}. However, the statistical
steady state found in the regime $-5/4<b<-1$ in \cite{94ABCFPV,96DM} is
close to thermodynamic 
equilibrium. This can be demonstrated via the properties of the
structure functions, defined by 
\begin{eqnarray}
S_2(k_n) &=& \langle|u_n|^2\rangle\ ,\label{eq:s2}\\
S_3(k_n) &=&  \text{Im} \{\langle u_{n-1}u_n u_{n+1}^* \rangle \} \ ,
\label{eq:s3}\\ S_4(k_n) &=& \langle|u_n|^4\rangle \ , \label{eq:s4}
\end{eqnarray}
etc. Indeed, in \cite{96DM} these objects were found in the inertial
range to be 
close to the exact solution in thermodynamic equilibrium which reads 
\begin{eqnarray}
S_2(k_{n}) &=& \frac{1}{B + A(a/c)^n}\ ,\label{s2sol}\\
S_3(k_n) &=&0 \ , \label{s3sol}\\
S_4(k_n)&=&S_2(k_n)^2 \ , \quad\text {etc.} \label{s4sol}
\end{eqnarray}
Formula (\ref{s2sol}) has two asymptotes: for small $n$ in agreement with
energy equipartition, and for large $n$ with enstrophy equipartition.
\begin{eqnarray}
S_2(k_n)&\sim& k_n^0\ ,\quad n_L\ll n\ll n_c \ , \label{S21}\\
S_2(k_n)&\sim& \left(\frac{c}{a}\right)^n\ , \quad n_c\ll n\ll n_f \ . 
\label{S22}
\end{eqnarray}
Here $n_c\approx\log(B/A)/\log(a/c)$ is the cross over shell separating the
two asymptotic scaling forms of $S_2$.  $A$ and 
$B$ are coefficients depending on the forcing and the dissipation. In particular, 
in this regime close to thermodynamic equilibrium, the cross over $n_c$ moves
to higher shells when the viscosity $\nu$ is reduced. Unless otherwise stated, we
choose parameters such that $n_c\ll  n_f$.
 
\begin{figure}
\centering
\includegraphics[width=.45\textwidth]{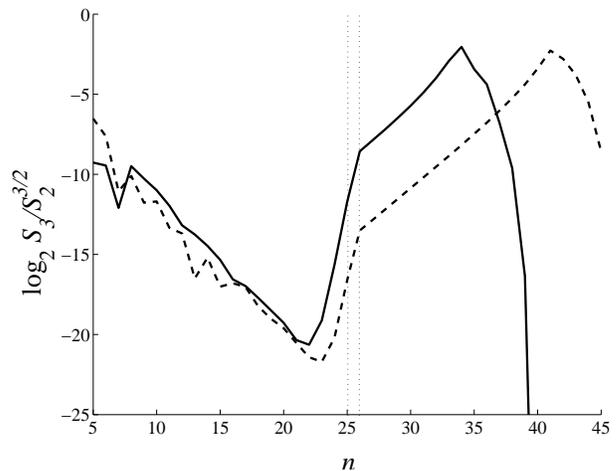}
\caption{The ratio $R(k_n)\equiv S_3(k_n)/S_2(k_n)^{3/2}$ as a
  function of $n$ for 
  $b=-1.1$ and $\nu=10^{-43}$ (dashed line) and
$\nu=10^{-33}$ (solid line). One clearly sees the three regimes predicted by
Eqs. (\ref{LEreg1}), (\ref{LEreg2}), (\ref{LEreg3}), with the minimum
occurring at $n=n_c$. The minimum deepens and moves to higher shells when the
viscosity is reduced. The vertical dotted lines designate the forcing scale.
Note that at small values of $n$ the statistics converges on much longer
time scales than
for large values of $n$; hence the remnant fluctuations around the scaling
prediction.}
\label{Fig1}
\end{figure}

Equation (\ref{s3sol}) implies zero fluxes. However in our simulations we
find in this regime a finite inverse flux of energy and a direct flux of enstrophy which do
not go to zero when $\gamma_n\to 0$. The fact that the fluxes do not vanish also implies
that $S_3$ is not exactly zero. 
One can write down the exact form of $S_3$, which is correct always
when there is a flux of energy or a flux of enstrophy~: 
\begin{eqnarray}
S_3(k_n)&\sim& k_n^{-1}\ , \: n_L\ll n\ll n_f \  
\mbox{(energy flux),} \label{S31}\\
S_3(k_n)&\sim& k_n^{-1}\left(\frac{c}{a}\right)^n\ , \: n_f\ll n\ll n_d\
\mbox{(enst. flux).}\
\label{S32}
\end{eqnarray}
A measure of the deviation of the statistics from Gaussian behavior 
is provided by the ratio
\begin{equation}
R(k_n)\equiv \frac{S_3(k_n)}{S_2(k_n)^{3/2}}\ ,
\label{defratio}
\end{equation}
which according to Eqs. (\ref{s2sol})-(\ref{S32}) has the three separate regimes
\begin{eqnarray}
R(k_n)&\sim& k_n^{-1}\ , \quad n_L\ll n\ll  n_c \ , \label{LEreg1}\\
R(k_n)&\sim&k_n^{-1}  \left(\frac{a}{c}\right)^{3n/2}\ , 
\quad n_c\ll n\ll  n_f \ , \label{LEreg2}\\
R(k_n)&\sim& k_n^{-1}  \left(\frac{a}{c}\right)^{n/2}\ ,
\quad n_f\ll n\ll n_d \ . \label{LEreg3}
\end{eqnarray}
These regimes are illustrated in Fig. \ref{Fig1}.

When $R$ is small, it provides a measure of the magnitude of the
fluxes compared to their standard deviations. $R$ is of order of
unity at the dissipative boundaries, while it reaches its minimal value at $n=n_c$. The
former follows from the fact that the dissipative boundaries are precisely
where the second order dissipative terms balance the third order transfer
terms. In fact the ratio $R$ cannot be larger
than unity whenever scaling prevails. One sees this directly
from the definitions (\ref{eq:s2}) and (\ref{eq:s3}):
\begin{equation}
S_3(k_n)/\sqrt{S_2(k_{n-1})S_2(k_n)S_2(k_{n+1}})\le 1\ .
\label{ratiolt1}
\end{equation}
Since $n_c$ moves to higher shells when the viscosity is reduced, the value
of $R$ at the minimum decreases: we divide a decreasing $S_3$ by an $S_2$ that
remains constant over a larger range of $n$. We thus
conclude that the quasi-equilibrium regime displays a alphabetical small parameter when
$\nu\to 0$. We will see that in the Kolmogorov regime there is only a numerical
small parameter. 

In ref. \cite{96DM} it was then discovered that there exists a transition
for $b$ crossing a critical value ($b=-5/4$ for $\lambda=2$) after which $S_2$ gains
a new form in the direct enstrophy flux regime, close to the
Kraichnan dimensional prediction \cite{67Kra}
\begin{equation}
S_2(k_n)\sim k_n^{-2[1+\log_\lambda(a/c)]/3} \ , \quad n_f\ll n\ll n_d\label{Kra} \ ,
\end{equation}
(up to small corrections).
We note that this prediction can be inferred from Eq. (\ref{LEreg3}) and
the condition (\ref{ratiolt1}). Indeed $R$ must be an {\em
  increasing} function of $n$ towards its small scale boundary, which yields
\begin{equation}
(a/c)\ge\lambda^2 \Leftrightarrow b\ge -a\left(1+\frac{1}{\lambda^2}\right)\ , 
\end{equation} 
or $b\ge -5/4$ for $\lambda=2$ and $a=1$. Thus for $b<-5/4$ Eq. (\ref{LEreg3})
can no longer be valid. While $S_3(k_n)$ does not change, $S_2(k_n)$
is replaced by the form (\ref{Kra}) and consequently Eq. (\ref{LEreg3}) is replaced
by
\begin{equation}
R(k_n)\sim k_n^0, \ n_f\ll n\ll n_d \ (b<-5/4)
\ . \label{enstreg3}
\end{equation}
\begin{figure}
\centering
\includegraphics[width=.45\textwidth]{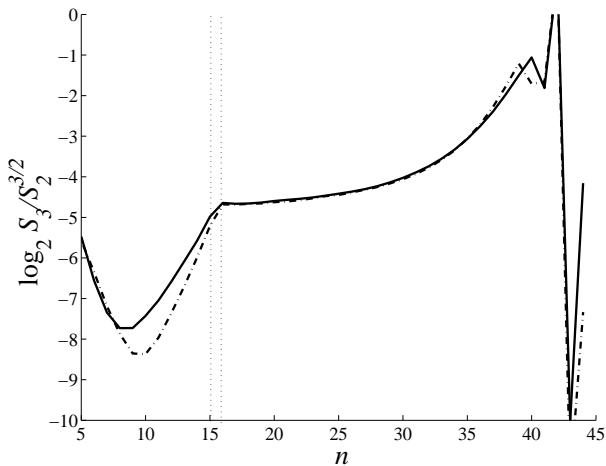}
\caption{Same as Fig. \ref{Fig1} but for 
  $b=-1.5$ (dot-dashed line) and $-1.6$ (solid line).
One clearly sees the three regimes predicted by
Eqs. (\ref{LEreg1}), (\ref{LEreg2}), (\ref{enstreg3}). Note 
that the constant regime $k_n^0$ is still effected by finite size effects
from the viscous end. However the two curves are essentially the same.}
\label{Fig2}
\end{figure}
In Fig. \ref{Fig2} we present this ratio as computed from numerical
simulations with the values of $b=-1.5$ and $-1.6$. We have used a total
of 46 shells, with $\nu=10^{-37}$, $\mu=10^{-3}$.
The forcing was on shells 15 and 16. The three
regimes are clearly seen, with the added important confirmation that this
ratio is of the order of unity at the two dissipative boundaries. 

Nevertheless previous work failed to find a similar phenomenon for the
range of scales that supports the inverse flux of energy. In that range the
statistics remained close to thermodynamic equilibrium, leading to
the common belief that shell models cannot be used to model 2-dimensional
turbulence.
We explain next that the statistical solution claimed for the regime
$b<-5/4$, i.~e. local thermodynamic equilibrium for the inverse flux of
energy and direct enstrophy cascade, predicts its own destruction when $b$
is reduced further  
beyond a critical value $b_c$ that we can compute analytically. Indeed the
set of Eqs. (\ref{LEreg1}), (\ref{LEreg2}), (\ref{enstreg3}) and the
condition (\ref{ratiolt1}) further
implies that $R$ cannot be a decreasing function of $n$ in the range $n_c\ll
n\ll n_f$, which implies 
\begin{equation}
\left(\frac{a}{c}\right)^{3/2} \ge \lambda \Leftrightarrow b\ge -a (1+\lambda^{-2/3})
 \ . \label{bc}
\end{equation}
Accordingly, for $b<b_c\equiv -a(1+\lambda^{-2/3})$ the quasi-equilibrium in the 
inverse energy flux regime can no longer be supported, and it changes
into a true cascade regime with K41 scaling. For
$\lambda=2$ this occurs at the critical value $b_c\approx -1.63$, where
$S_2(k_n)$ assumes the scaling form
\begin{equation}
S_2(k_n) \sim k_n^{-2/3}\ ,\ n_L\ll n\ll n_f \ .
\label{s2invc}
\end{equation} 
Note that Eq. (\ref{s2invc}) implies the collapse $n_c\to n_L$. 
\begin{figure}
\centering
\includegraphics[width=.45\textwidth]{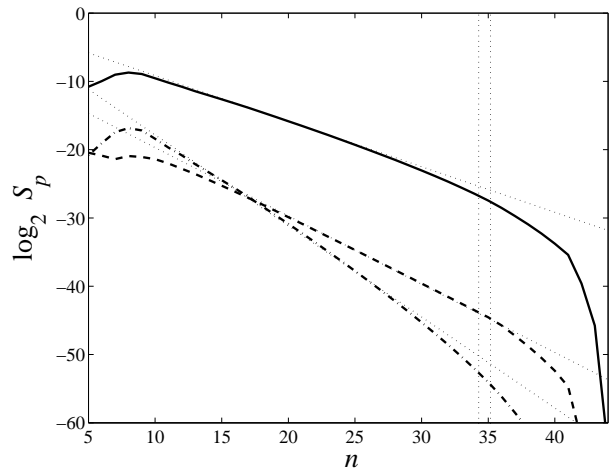}
\caption{Second (solid line), third (dashed line) and fourth (dot-dashed
line) order structure functions in the inverse cascade regime for $b=-1.9$. 
The vertical dotted lines indicate the forcing range
$n_f=35,36$. The dotted lines have K41 slopes of -2/3, -1 and 
and -4/3 respectively.}
\label{Fig3}
\end{figure}
In Fig. \ref{Fig3} we show the results of simulations at $b=-1.9$,
with forcing at shells $n_f=35,\:36$ and otherwise the same parameter
values as in Fig. \ref{Fig1}. The agreement with K41 scaling is apparent.
We note that the scaling laws (\ref{s2invc}) and (\ref{S31}) (which 
remains true in this regime) imply that $R(k_n)$ becomes constant
as a function of $k_n$. Thus we cannot display an alphabetical small
parameter anymore. Nevertheless, the measurement of the constant
value of $R$ in the inverse cascade regime yields a number of the
order of 0.02 or less. We thus have a {\em numerical} small parameter,
that is similar in magnitude to the corresponding value of $R$ in
2-dimensional turbulence \cite{00BCV}.

In summary, we exhibited a new regime of the statistical properties of
shell models in which inverse energy cascade exists side by side
with a direct enstrophy cascade. The statistical objects satisfy
scaling laws in close correspondence with the Kraichnan dimensional
predictions for 2-dimensional turbulence. Since this model is so much
simpler than 2-dimensional Navier-Stokes equations, it should provide
useful grounds to understand the phenomenon theoretically. Such a discussion
and a more detailed account of our numerical findings will be presented
elsewhere \cite{02GLPP}.

\acknowledgments
This work has been supported in part by the European Commission
under a TMR grant, the German Israeli Foundation, and the
Naftali and Anna Backenroth-Bronicki Fund for Research in
Chaos and Complexity. TG thanks the Israeli Council for 
Higher Education and the Feinberg 
postdoctoral Fellowships program at the WIS for financial support.

\end{document}